\documentstyle[11pt,epsf,newpasp,twoside]{article}

\def\etal{et al }

\def\msun{M_\odot}
\def\nbody{$N$-body\ }

\def\edcomment#1{\iffalse\marginpar{\raggedright\sl#1\/}\else\relax\fi}
\reversemarginpar

\begin{document}
\title{The Gravitational Million-Body Problem}
 \author{Douglas C. Heggie}
\affil{University of Edinburgh, Department of Mathematics and
 Statistics, King's Buildings, Edinburgh EH9 3JZ, UK}

\begin{abstract}
We review what has been learned recently using \nbody simulations about
the evolution of globular clusters.  While simulations of star
clusters have become more realistic, and now include the evolution of
single and binary stars, the prospect of reaching large enough $N$ is
still a distant one.  Nevertheless more restricted kinds of
simulations have recently brought valuable progress for certain
problems of current observational interest, including the origin and
structure of tidal tails of globular clusters.  In addition, such
simulations have forced us to rethink some basic aspects of stellar
dynamics, including, in particular, the process of escape.  Finally we
turn to faster, approximate  methods for studying star cluster
dynamics, where the role of \nbody simulations is one of calibration.
\end{abstract}

\section{Introduction}

The title of this review unfortunately disguises the fact that it is
about globular clusters, in the traditional sense.  Actually, ``a
million'' is somewhat on the large side in this context.  Though the
total number of stars in a globular cluster is not known, it may be
estimated roughly by converting the luminosity (Harris 1996) to the
mass, using the method of Mandushev, Staneva, \& Spasova (1991), and
adopting a mean stellar mass of $0.34M_{\sun}$, which corresponds
approximately to a power-law mass function with slope $-1$ (Kroupa
2001) between turnoff and the hydrogen burning limit.  The result is
(Fig.1) that the median value is about $3\times10^5$, and roughly 10\%
of all clusters have more than one million stars.  In this review we
highlight some of the ways in which $N$-body simulations are being
used in studying the dynamical theory of globular star clusters.

\section{Comprehensive Calculations\footnote{At Tokyo several people
expressed some distaste for the phrase ``kitchen-sink calculations'',
which I used in my talk.}}

\begin{figure}
\plotfiddle{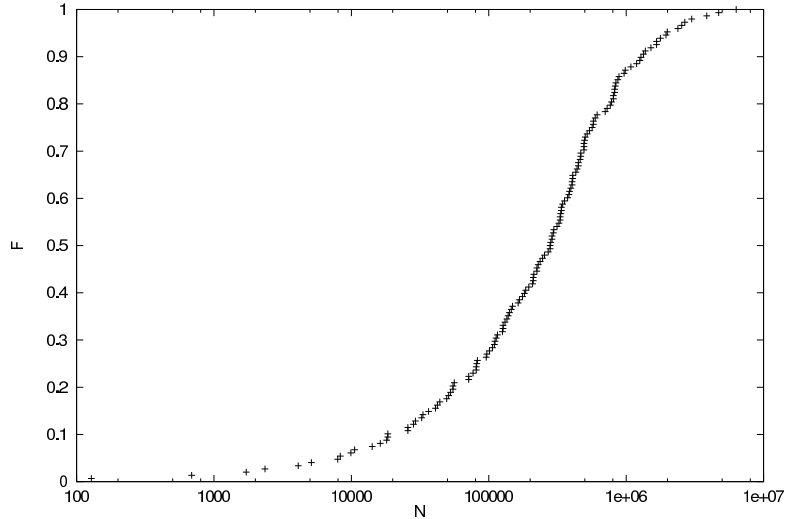}{2.5truein}{-90}{40}{40}{-170}{210}
\caption{Cumulative distribution of approximate number of stars in the
clusters of the galactic globular cluster system, derived as in the
text.}
\end{figure}

No simulation of a globular cluster will ever be complete, but rapid
strides are being made to extend the traditional techniques of stellar
dynamics so as to include stellar evolution.  As a result there is a
growing body of simulations with the following typical characteristics:
(i) 1 star is represented by 1 particle (``star-by-star
calculation''); 
(ii) forces include star-star gravity, a galactic tide (for a cluster
on either a circular or elliptic galactic orbit), and disk shocking;
(iii) treatment of stellar collisions;
(iv) inclusion of stellar properties (luminosity, colour,
metallicity) and stellar evolution; and
(v) inclusion of (primordial) binaries and binary evolution.

There is a strong synergy between the development of such codes and
the development of special-purpose hardware (see the papers by
Aarseth, Makino and the ``starlab'' group in this volume.)  Table 1
lists some recent comprehensive simulations, where $f_{bin}$ is the
initial fraction of binaries.
The scientific value of such work is well illustrated by Hurley et al
(2001), who quantified the role of dynamical influences on the
population of blue stragglers in an old open cluster.  Such a result
has been a long standing goal of this programme of research.

\begin{table}
\begin{center}
\caption{		    Some comprehensive calculations}
\end{center}
\begin{tabular}{llrr}
\tableline
Author		&		Object&	      $N(0)$&	$f_{bin}$\\
\tableline
Hurley et al (2001)&		M67&		15000&	50\%\\
Portegies Zwart et al (2001a)	&Pleiades, etc& 3000&	50\%\\
Portegies Zwart et al (2001b)   &Arches, etc  & 12000&	0\%\\
Kroupa et al (2001)*            & ``Pleiades"    &10000&	100\%\\
\tableline\tableline
*No Roche Lobe mass transfer, etc 

\end{tabular}

\end{table}

Despite such successes, it is disappointing that not one of these is a
simulation of a globular cluster, in the traditional sense of the
term.  Even with GRAPE-4 it was possible to study a model with $N =
32768$ well past core collapse (Makino 1996), which includes at least
the smallest 10\% of the existing galactic globular clusters.  

There are two reasons why simulation of any globular cluster is much
harder than is suggested by the particle number alone.  One is the
requirement for a substantial population of primordial binaries.
Though it has been argued that the computational load associated with
binaries becomes negligible when $N$ is sufficiently large (Makino \&
Hut 1990), this asymptotic limit is not yet in sight.  The second
reason is that the present particle number in a globular cluster may
be a poor guide to what is required.  Hurley et al (2001) estimated
that even M67 had 7 times as many stars at birth as at present, and were
therefore unable to simulate its early evolution.  The present day
mass function of the globular cluster NGC 6712 can be understood if it
has lost about 99\% of its original mass (Takahashi \& Portegies Zwart
2000), and since its current mass is estimated to be of order
$10^5\msun$ (Pryor \& Meylan 1993), its likely original particle
number puts it far beyond reach.

One conclusion from this discussion is that, if a comprehensive
simulation of a globular cluster is carried out within the next few
years, it will tell us only about the evolution of the few clusters
(if there are any) which are of low mass now and always have been.
Study of the dynamical evolution of globular clusters cannot wait
until such simulations can cope with all clusters.

One of the aims of comprehensive simulations, as we have seen, is the
study of the interaction between stellar and dynamical evolution, and
the effects of collisions.  While the dynamics of point masses can be
computed with precision, the effect of collisions is very uncertain.
As Sills \etal (2001) make clear, the effects of rotation on a merger
remnant are dramatic.  Since this is likely to be the largest source
of uncertainty in studying the effects of collisions in
globular clusters, it may be perfectly acceptable to adopt a
model in which the point-mass dynamics is treated approximately, such
as a Monte Carlo model (Sec.5).  These are much faster than direct
$N$-body methods, and can reach much larger ``particle'' numbers,
ensuring that rare (but observable) species are not neglected.

\section{			Restricted Simulations}

\begin{figure}
\plotfiddle{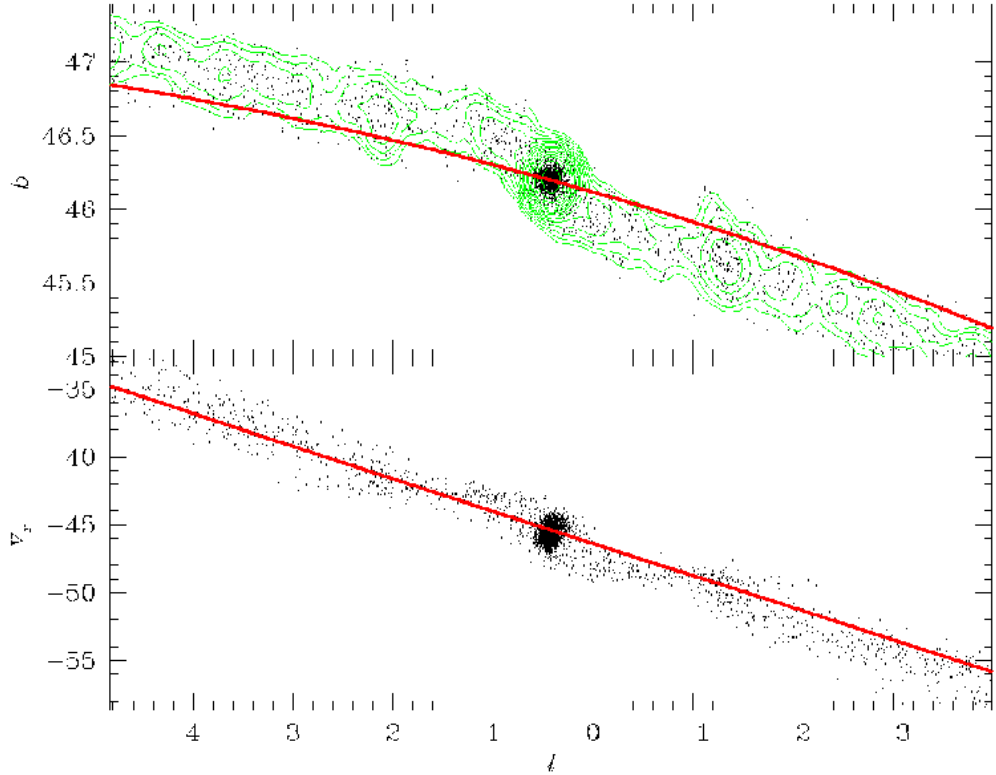}{2.3truein}{-90}{50}{50}{-225}{290}
\caption{Simulation of Pal 5 (Dehnen, pers. comm., reproduced by kind
permission of the
author).  The orbit chosen is consistent with the present position,
radial velocity and proper motion, and used the galactic potential of model 3 in Dehnen \&
Binney (1998).  The initial condition is a King model with $N =
10^4$.  Upper panel: position on the sky ($l,b$); lower panel: radial velocity
against longitude.  Note the clumps in the tail in the upper panel;
they resemble those in Pal 5
itself (Odenkirchen et al 2001, where they are attributed to recent
disk shocks).}
\end{figure}

Considerable progress has been made in recent years in understanding
important aspects of the dynamics of globular clusters without
attempting a comprehensive simulation.  The example we shall
concentrate on here is the simulation of tidal tails.  It is a
theoretical problem that has been stimulated by observations
(Grillmair et al 1995, 1996; Kharchenko, Scholz, \& Lehmann 1997; Leon,
Meylan, \& Combes 2000; Testa \etal 2000; Odenkirchen et al 2001). For
this purpose one need include only (i) point mass dynamics, (ii)
relaxation, (iii) the galactic tide, and (iv) the galactic orbit.
Binaries and stellar evolution can be ignored.  This is a pure stellar
dynamics problem, and several sets of simulations have been carried
out:  Moore (1996); Combes, Leon \& Meylan (1999), Murali \& Dubinski
(1999), Johnston, Sigurdsson \& Hernquist (1999), Dehnen
(pers. comm., and Fig.2), and others.  The list would be
much longer if one added comparable work on satellite galaxies.

What is not always clear from simulations is the mechanism by which
stars escape from a cluster into the tidal tail.   The most obvious
possibilities are time-dependent tides (disk and bulge shocking) and
relaxation.  But such simulations usually cover only one or two Gyr of
evolution, and it is possible that the choice of initial conditions
plays a role, just as in the case of circular orbits (Fukushige \&
Heggie 2000).  The other hard thing to get right is relaxation, though
the simulations of Combes et al (1999) show that this can be done.

The role of these various mechanisms can be considerably clarified by
a judicious choice of initial conditions and parameters.  Baumgardt
(this volume) begins at apogalacticon, with a cluster whose radius
corresponds to the Roche radius at perigalacticon.  The rationale for
this is {\sl not} the assumption that, over the course of many
galactic orbits, the radius of a cluster is set by its Roche radius at
perigalacticon; such an assumption would be in contradiction with
observational evidence (Ninkovic 1985, Allen \& Martos 1988, Meziane
\& Colin 1996; Brosche, Odenkirchen \& Geffert 1999).  Rather, if the
initial radius of the cluster is either much greater or much smaller
than the perigalactic tidal radius, then it is clear that initial
conditions have much to do with the subsequent mass loss.

Baumgardt's finding from his \nbody simulations is that, if one
estimates the lifetime on the basis of mass lost in the first few
orbits, as is often done, one finds that the lifetime increases with
$N$ roughly as the relaxation time, until sufficiently large $N$ is
reached, and then the lifetime is approximately independent of $N$.
This is what would be expected if relaxation dominates for small $N$
and tidal heating dominates for large $N$.  If, however, one waits for
as many galactic orbits as are required to lose 
a certain fraction (such as 25\%) of the mass, one finds that the mass
loss is dominated by relaxation, for all $N$ up to the largest values
he was able to study.  Though this $N$ is still much smaller than the
number of stars in the larger globular clusters, Baumgardt's $N$-body
models used particles of equal mass, and their relaxation time
actually corresponds to the relaxation time of a much larger cluster
with unequal masses.

One draws two conclusions from this study.  One is that the initial
conditions matter; it is only after the lapse of a sufficient number
of galactic orbits that the mass loss rate reaches its asymptotic
size.  Second, the escape rate is dominated by relaxation, and not by
bulge shocking, even on a very eccentric galactic orbit.  It remains
to be seen, however, whether {\sl disk} shocking is ever important.

\section{			 New Stellar Dynamics}

While it is clear that there are quantitative questions which can only
be answered from \nbody simulations, has anything qualitatively new
come out of them?  Indeed, yes, but the list is not a long one.  Most
recently, we have learned that the lifetime of clusters does not scale
with $N$ in the expected manner, and the present section summarises
this discovery.

The aim of a collaborative experiment carried out in 1997 (Heggie et
al 1998; Heggie, this vol.) was to compare results of various kinds of
codes (and not only \nbody codes) in studying the dynamical evolution
of a specific object.  Since the number of particles was of order $250
000$, it was necessary to scale \nbody simulations by $N$.  When this
was done by the relaxation time, it was found that the lifetime of the
system depended on $N$, though its internal evolution (e.g. time to
core collapse) was fairly insensitive to $N$.

Fukushige suggested (pers. comm.) that the problem lay in the time
taken for stars to escape, once they had enough energy to do so.  If,
however, it was assumed that the escape time scale is proportional to
the crossing time, $t_{cr}$, then it seemed hard to understand how the problem
persisted for the large values of $N$ (up to 64K) for which results
were available.  The lifetime of a cluster should vary nearly with the
relaxation time, except for $N$ up to a few hundred, as in the escape
theory of King (1959).  Fukushige \& Heggie (2000), however, found
that the time taken to escape is a sensitive function of the energy, $E$, of
a star:  $t_{esc}\propto (E - E_{esc})^{-2}$, where $E_{esc}$ is the
energy needed to escape (i.e. the potential at the Lagrange point,
where the potential well of the cluster opens out into the rest of the
galaxy). 

The next question was to find the appropriate energy to use in this
formula.  The answer is provided in Baumgardt (2001a).  (His argument
is more detailed than the following, but leads to the same scalings.)
Unescaped stars with energies just above $E_{esc}$ are relaxing during
the time they take to escape, and so their excess energy is $E -
E_{esc} \propto (t_{esc} / t_{rlx})^{1/2}$, since relaxation is a
diffusion process.  (Here, $t_{rlx}$ is the relaxation time.)  
By solving these two relations we find the
following conclusions:
(i) $t_{esc}\sim \left(t_{cr} t_{rlx}\right)^{1/2}$, much longer
than the crossing time, and quite different
from previous expectations;  this helps to explain why the effects of
the escape time persist for much larger $N$ than expected;
(ii) $E - E_{esc}\propto  t_{rlx}^{-1/4}$;  this is also the fraction
of stars inside the cluster with energy above $E_{esc}$;  this
fraction is still a few percent, even for systems of the size of a
typical globular cluster; and
(iii) from the previous two conclusions, the time to lose some
fraction (say, half) of the mass of the cluster is $t_M\propto
t_{rlx}^{3/4}$.

The last is the most far-reaching conclusion.  Though the theory
applies only for circular galactic orbits, Baumgardt has shown
empirically (these proceedings) that the same conclusion appears to
hold for clusters on elliptic galactic orbits.  And yet Vesperini \&
Heggie (1997) studied the evolution of the mass function of clusters
assuming that $t_M\propto t_{rlx}$, which is now known to be wrong.
Indeed a great deal of recent modelling of globular cluster systems
(e.g. Capuzzo-Dolcetta \& Tesseri 1997, Gnedin \& Ostriker 1997,
Murali \& Weinberg 1997, and other work of authors already quoted) may
require revision.  In effect, the new scaling of $t_M$ means that the
lifetime of a globular cluster through dynamical processes is smaller
(for large $N$) than one would have thought.  Baumgardt (2001b) shows
that one consequence of this is that the initial conditions of the
cluster system are more completely erased.

\section{		    Calibration of Faster Methods}

\subsection{Faster methods for star cluster dynamics}

\begin{figure}
\plotfiddle{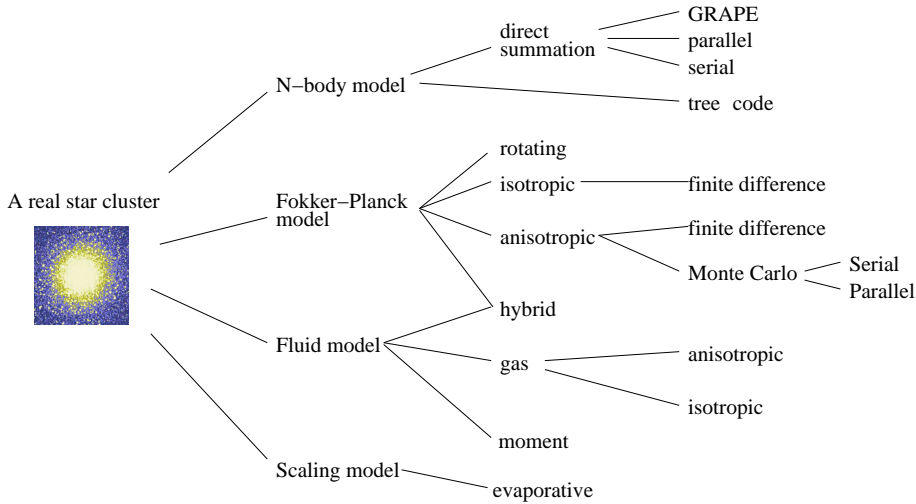}{2.3truein}{0}{75}{75}{-180}{-10}
\caption{ Models of the dynamics of dense stellar
systems.  We do not include all hybrids, or methods used
in other areas of stellar dynamics.}
\end{figure}

Besides \nbody simulations, there are several other methods for
studying the dynamical evolution of globular clusters (Fig.3).  There are
scaling methods, in which the cluster is characterised by nothing more
than its mass and length scale (for example), Fokker-Planck methods,
gas methods, and various hybrids.  Even tree methods may function well
for some important problems (McMillan \& Aarseth 1993, Arabadjis \&
Richstone 1998, Miocchi \& Capuzzo-Dolcetta\ 2001).  All these methods
are faster and involve more approximations, and $N$-body simulations
are required in order to calibrate them.

We have already seen
(Sec.2) that such methods may be quite appropriate for problems where the
accuracy of the stellar dynamics is not the main uncertainty.  Fast
methods are also appropriate for the following kinds of problems:
\begin{enumerate}
\item{\sl Evolution of cluster systems:}  For references see Sec.4.
For this purpose it must be possible to study the dynamical evolution of many
globular cluster models with the full range of realistic $N$.
At present this is possible only with fast approximate methods.
\item {\sl Modelling individual objects:} Fokker-Planck and
(occasionally) fluid methods have been used to construct evolutionary
models for comparison with observations of specific clusters,
including M3 (Angeletti, Dolcetta \& Giannone 1980), M71 (Drukier, Fahlman \& Richer 1992), NGC 6624 (Grabhorn et al 1992), NGC 6397 (Drukier 1993,
1995), M15 (Grabhorn et al 1992, Phinney 1993, Dull et al 1997) and M30
(Howell, Guhathakurta \& Tan 2000).  This requires extensive trial and
error, with variations of the initial conditions of the model, and
fast methods are essential.  
One motivation for such work has been the
need to select suitable observational fields in clusters such that the
local mass function requires least correction to the global mass function.
\item{\sl Predicting initial conditions:} This is the theorists'
counterpart of the previous point.  If one wants to construct an
$N$-body model of a given object, considerable trial and error is
needed in order to find appropriate initial conditions.  Faster
methods should be able to give a rapid but approximate answer, which
can then be refined with a  minimal number of \nbody simulations.
\end{enumerate}

\subsection{Recent examples}

\subsubsection{Scaling models}

As part of a study of the dynamical evolution of the mass function of
a star cluster, Vesperini \& Heggie (1997) gave a simple formula for
the evolution of the mass, $M(t)$, i.e. $M(t) = M(0) - \Delta M_{s.e.}
- 0.828M(0)t/F_{CW}$, where $\Delta M_{s.e.}$ is the contribution from
stellar evolution (which also depends on the mass function), and
$F_{CW}$ is the ``family'' parameter of Chernoff \& Weinberg (1990).
For a given mass function, it is proportional to the relaxation time
for average conditions inside the tidal radius, and is defined by
$F_{CW} = (M/\msun)(R_g/{\rm kpc})(220kms^{-1}/v_g)\gamma^{3/2}/\ln
N$, where $M$ is the (original) mass of the cluster, moving on an
orbit of radius $R_g$ at speed $v_g$.  $\gamma$ depends on the
distribution of mass in the galaxy, is unity for a point-mass galaxy,
and $(3/2)^{1/3}$ for an isothermal potential.

The above formula for $M(t)$ has been used by Vesperini (1998, 2000,
2001) in studies of the evolution of cluster systems.  Its relevance
to the present review is that it was based on the simplest scaling
considerations, but the numerical coefficient was obtained by
examination of $N$-body simulations.  In other words, $N$-body results
were used to calibrate a simpler theoretical model.

We now briefly consider the lifetime of a cluster.  In the absence of
stellar evolution, the above formula gives a cluster lifetime $t_{VH}
= F_{CW}/0.828$ in Myr.  The coefficient depends weakly on the
structure of the initial model.  
A similar formula for $M(t)$ has been given by Portegies Zwart et al
(2001b), which leads to a lifetime 
$t_{PZMMH} = 0.29t_{rxt}$.  Here $t_{rxt}$ is another measure of the
relaxation time within the tidal radius, defined by $t_{rxt} =
0.138M^{1/2}r_t^{3/2}/(\bar mG^{1/2}\log_{10}\Lambda)$ (cf. Spitzer
1987, eq.(2-63), except that Spitzer uses the natural logarithm in the
Coulomb factor $\log\Lambda$).  In this formula, $\bar m$ is the mean
stellar mass.  

For the galactic mass law used by Portegies Zwart et al, $\gamma =
(3/1.8)^{1/3}$, and so it is easy to show, if we equate the two forms
of the Coulomb logarithm, that the ratio of the lifetimes is
$t_{VH}/t_{PZMMH} \simeq 0.34\msun/\bar m$.  The dependence on $\bar
m$ arises because $F_{CW}/t_{rxt}$ also depends on $\bar m$.  If it is
assumed that the lifetime is proportional to $t_{rxt}$ (independent of
the mass function, cf. de la Fuente Marcos 1995), then it is
necessary to modify $t_{VH}$ to $t_{VH} = (F_{CW}/0.828)(\bar
m_{VH}/\bar m)$, where $\bar m_{VH}$ is the value in the simulations
by Vesperini \& Heggie (1997).  In fact for most of their work, which
used a mass function $f(m)\propto m^{-2.5}$ for $0.1\msun<m<15\msun$
we have $\bar m_{VH}\simeq 0.28\msun$.  The effect of this change
(which is also necessary in the formula for $M(t)$) is to bring the
two independent estimates of cluster lifetime into fair agreement.  It
must be borne in mind, however, that the assumption that the lifetime
is proportional to the relaxation time is incorrect (Sec.4), and that
$\bar m$ is time-dependent.

Despite the fact that lifetime (in the absence of stellar evolution)
does not depend much on the initial structure (if the cluster 
initially fills its Roche lobe), it would be useful to have an equally
simple way of estimating the time-dependence of the concentration
parameter.

\subsubsection{Fokker-Planck models}

The comparison between $N$-body and Fokker-Planck models (computed
with finite differences) has a considerable history, but recent work
appears to have brought the two methods into reasonable agreement,
with one caveat, discussed below.  The central issue is the escape
criterion, as shown by Takahashi \& Portegies Zwart (1998, 2000).
These papers show how a free parameter in the Fokker-Planck
specification can be determined by calibration using the results of
\nbody models.  Incidentally, the second of these is also an extensive
survey of globular cluster models, which essentially supersedes the
comparable extensive Fokker-Planck work of Chernoff \& Weinberg
(1990).

As Takahashi \& Portegies Zwart themselves point out, the caveat in
this is that the \nbody models used in this survey adopt a tidal
cutoff.  Since it is known that the lifetime varies with $N$ in a
different way when a tidal field is used (Sec.4), it remains to be
seen how well the agreement between Fokker-Planck and \nbody models
survives in the more realistic case.

\subsubsection{Monte Carlo models}

Though these can be regarded also as Fokker-Planck models, they offer
some advantages over finite-difference models, and have received
increasing attention in recent years.  While part of this effort
involves calibration with \nbody models, the real purpose of this
little subsection is to say something about the present state of play.

The largest models to date are those of Freitag \& Benz (2001), who present
results of Monte Carlo simulations with up to $2\times10^6$ shells.
At such levels it becomes possible to represent many globular clusters
with one shell per star.  These models, which are motivated by
problems of galactic
nuclei, do not include binaries, but
have been extended to include stellar collisions (Fig.4).  For
models including the effects of dynamically formed binaries the record
is held, at time of writing, by Giersz (this volume).  Reaching well
past core collapse with $N=10^6$ shells, as he does, is a notable landmark.
Primordial binaries are excluded from these models, but a 10\%
population of primordial binaries are present in computations with up
to $3\times10^5$ shells by Rasio's group (Rasio, Fregeau \& Joshi,
2001; Fig.5).  Soon
they will compute few-body interactions ``on the fly'', instead of
using cross sections.  All these developments give great promise for
the Monte Carlo method in the coming years.

\begin{figure}
\plotfiddle{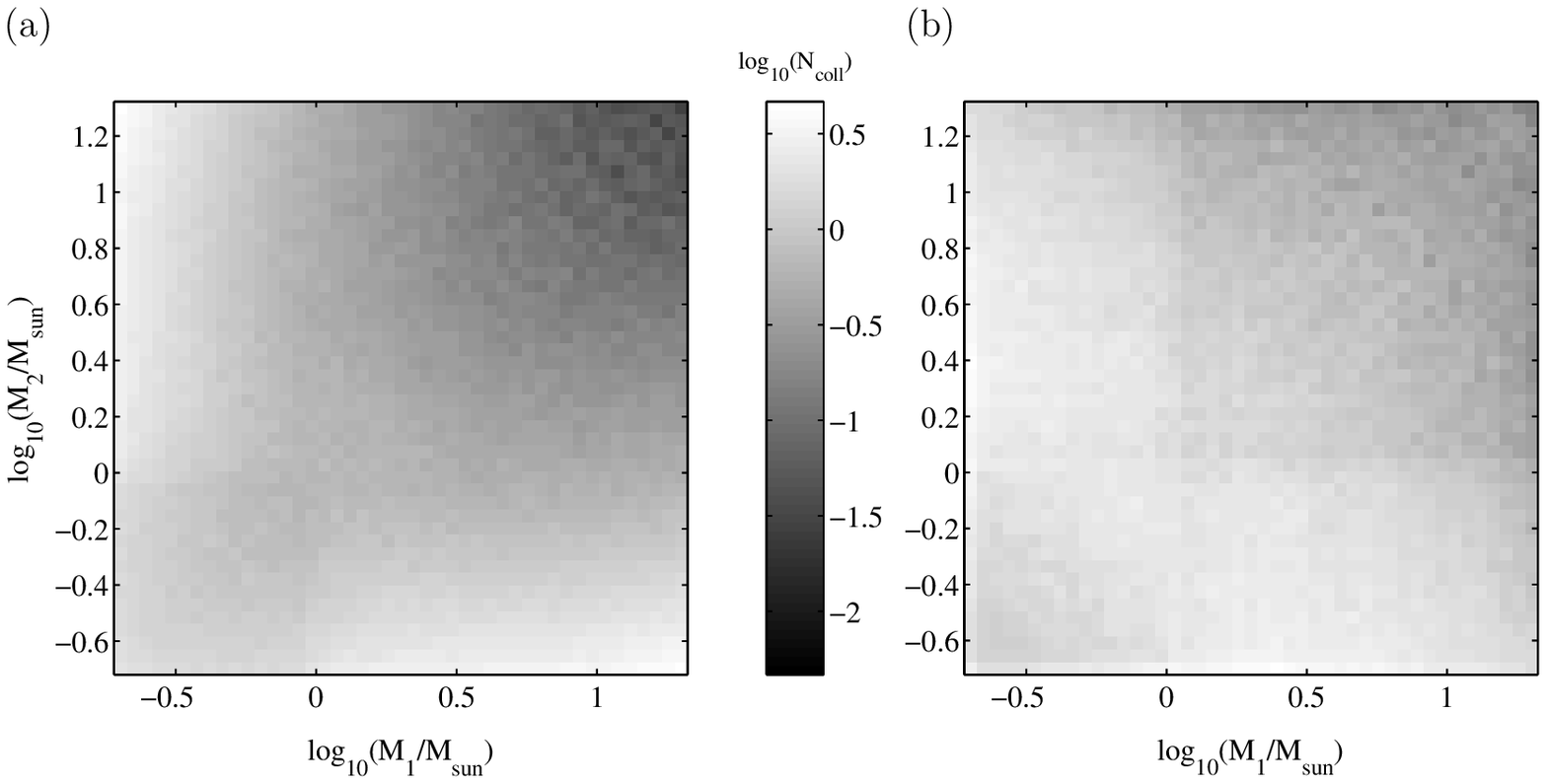}{2.5truein}{0}{70}{70}{-200}{-200}
\caption{Comparison of the collision rate between stars of mass $M_1$
and $M_2$ in a static Plummer model
(left) and in a
Monte Carlo simulation (right) with $512000$ shells, but without
stellar evolution (from Freitag 2000, with kind permission).  The collision rate among more massive stars is
greatly enhanced by mass segregation.}

\end{figure}

\begin{figure}
\plotfiddle{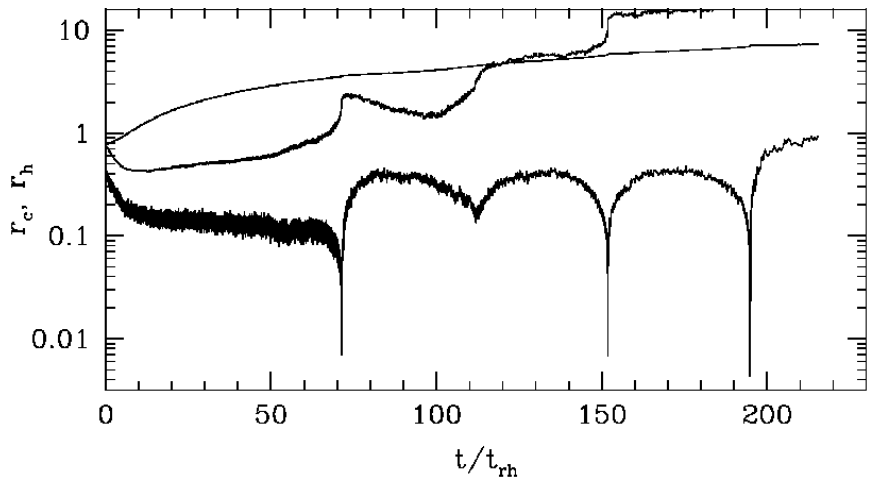}{2.5truein}{-90}{40}{40}{-170}{225}
\caption{Simulation of an isolated system with $N=3\times10^5$
initially, including 10\% primordial binaries (F. Rasio, by kind
permission).  Radii plotted against time are, from the bottom: core
radius, half-mass radius of binaries, half-mass radius of all stars.
Core collapse is avoided for about 70 initial half-mass relaxation
times, when gravothermal oscillations begin.}

\end{figure}

\section{Conclusions}

This rather selective review has concentrated on what \nbody models
are telling us about the dynamics of globular star clusters.
Comprehensive simulations, which include details of evolution of single and
binary stars, are at an exciting stage, but are still limited to rich
open clusters.  It seems unlikely that comprehensive simulations of
globular clusters will be possible in the near future.  Much progress
can be made on some interesting kinds of dynamical problems about
globular clusters (e.g. tidal tails) using more restricted kinds of
simulation, which are perfectly feasible now.  One of the dynamical
lessons that has been learned in recent years is that it is vital to
implement a correct treatment of tidal effects, even in the case of a
steady tide (i.e. a circular galactic orbit).  Monte Carlo methods now
have the dynamical aspects of globular clusters within reach.  When a
detailed treatment of stellar evolution has been incorporated (at the
level which has already been successful in the best \nbody codes),
these codes may well prove optimal for the comprehensive study of
globular star clusters, until the time when real \nbody models are
fast enough.

\acknowledgements I thank the organisers of the meeting for their
support, and I am grateful for comments from H. Baumgardt,
S.P.F. Portegies Zwart and E.  Vesperini on scientific matters.  I
also thank W. Dehnen, M. Giersz, M. Freitag and F. Rasio for the
provision of unpublished results.  The picture of M15 in Fig.3 was
obtained using NASA's {\sl SkyView} facility
(http://skyview.gsfc.nasa.gov) located at NASA Goddard Space Flight
Center.

\end{document}